\documentclass[conference]{IEEEtran}

\usepackage{epsfig}
\usepackage{times}
\usepackage{float}
\usepackage{afterpage}
\usepackage{amsmath}
\usepackage{amstext}
\usepackage{amssymb,bm}
\usepackage{latexsym}
\usepackage{color}
\usepackage{graphicx}
\usepackage{amsmath}
\usepackage{amsthm}
\usepackage{graphicx}
\usepackage[center]{caption}
\usepackage{pstricks}
\usepackage{subfigure}
\usepackage{booktabs}

\def \eu{\mathrm{e}}

\newcommand{\snr}{{\small\mathsf{SNR}}}
\newcommand{\inr}{{\small\mathsf{INR}}}
\newcommand{\mud}{{\small\mathsf{I_d}}}
\newcommand{\mug}{{\small\mathsf{I_g}}}

\newtheorem{thm}{Theorem}
\newtheorem{cor}[thm]{Corollary}
\newtheorem{lem}[thm]{Lemma}

\title{On Discrete Alphabets for the Two-user Gaussian Interference Channel with One Receiver Lacking Knowledge of the Interfering Codebook} 
\author{
\IEEEauthorblockN{Alex Dytso, Daniela Tuninetti, and Natasha Devroye\\}
\IEEEauthorblockA{%
University of Illinois at Chicago, Chicago IL 60607, USA,\\
Email: {\tt odytso2, danielat, devroye @ uic.edu}}%
}

\begin{document}

\maketitle

\begin{abstract}
In multi-user information theory it is often assumed that every node in the network possesses all codebooks used in the network. This assumption is however impractical in distributed ad-hoc and cognitive networks. 
This work considers the two-user Gaussian Interference Channel with one {\em Oblivious Receiver} (G-IC-OR), i.e., one receiver lacks knowledge of the interfering cookbook while the other receiver knows both codebooks.
We ask whether, and if so how much, the channel capacity of the G-IC-OR is reduced compared to that of the classical G-IC where both receivers know all codebooks. Intuitively, the oblivious receiver should not be able to jointly decode its intended message along with the unintended interfering message whose codebook is unavailable.
We demonstrate that in strong and very strong interference, where joint decoding is capacity achieving for the classical G-IC, lack of codebook knowledge does not reduce performance in terms of generalized degrees of freedom (gDoF). 
Moreover, we show that the sum-capacity of the symmetric G-IC-OR
is to within $O(\log(\log(\snr)))$ of that of the classical G-IC. 
The key novelty of the proposed achievable scheme is the use of a discrete input alphabet for the non-oblivious transmitter, whose cardinality is appropriately chosen as a function of $\snr$. 
\end{abstract}

\section{Introduction}
A classical assumption in multi-user information theory is that each node in the network possesses knowledge of the codebooks used by every other node.  However, such assumptions might not be practical in heterogeneous, cognitive, distributed or dynamic networks 
For example, in very large  ad-hoc networks, where nodes enter and leave at will, it might not be a practical assumption that new nodes learn the codebooks of old nodes and vice-versa. On the other hand, in cognitive radio scenarios, where new cognitive systems coexist with legacy systems, requiring the legacy system to know the codebook of the new cognitive system might not be viable. 
This motivates the study of networks where each node possesses only a subset of the codebooks used in the network. We will refer to such systems as networks with \emph{partial codebook knowledge} and to nodes with only knowledge of a subset of the codebooks as \emph{oblivious receivers}. 

\subsection{Past Work}
To the best of our knowledge systems with partial codebook knowledge were first introduced in \cite{sand_decentr_proces}. In \cite{sand_decentr_proces} lack codebook knowledge was modeled by using {\it codebook indices}, which index the random encoding functions that map the messages to the codewords. If a node has codebook knowledge it knows the index (or instance) of the random encoding function used; else it does not and the codewords essentially look like the symbols were produced in an independent, identically distributed (i.i.d.) fashion from a given distribution. 
In \cite{simion_codebook} and \cite{CodebookYEner} this concept of 
partial codebook knowledge was extended to model {\it oblivious relays}, where only multi-letter capacity expressions were obtained.  
As pointed out in \cite[Section III.A]{simion_codebook} and \cite[Remark 5]{CodebookYEner}, these capacity bounds are ``non-computable'' in the sense that it is not known how to find the optimal input distribution in general. In particular, the capacity achieving distribution  for the practically relevant Gaussian noise channel remains an open problem. 

In \cite{oneObliviISIT2013} we introduced the two-user Interference Channel (IC) with one Oblivious Receiver, referred to as the IC-OR. In the IC-OR, one receiver has full codebook knowledge (as in the classical IC), but the other receiver only has partial codebook knowledge (it knows the codebook of its desired message, but not that of the interfering message). The capacity region of the IC-OR was characterized to within a constant gap for the class of {\it injective semi-deterministic} IC in the spirit of~\cite{telatar2008bounds}. In particular, the capacity of the real-valued Gaussian IC-OR  (G-IC-OR) was characterized to within 1/2~bit per channel use per user; however, the input distribution achieving such a gap was not found. 
In \cite[Section V.B]{oneObliviISIT2013} it was remarked that a carefully chosen i.i.d. Pulse Amplitude Modulation (PAM) can outperform i.i.d. Gaussian inputs for the given achievable rate region expression, and it was thus 
conjectured that discrete inputs may outperform Gaussian signaling in the strong and very strong interference regimes.

\subsection{Contributions and Paper Outline} 
After formally introducing the IC-OR in Section \ref{sec:channelModel}, we show our main contributions:
\begin{enumerate}

\item 
In Section \ref{sec:MainTool} we introduce a new lower bound on the mutual information achievable by a discrete input on a point-to-point Gaussian noise channel, which will serve as the main tool in the derivation of our achievable rate region for the G-IC-OR.

\item 
To understand the utility of this new tool, in Section \ref{sec:ptpExample} we show 
how to choose the cardinality of the discrete input in a point-to-point Gaussian noise channel such that 
the rate achieved is to within $O\left(\log\left(\log(\snr)\right)\right)$ of the (in this case known) capacity.
This in turn shows that a discrete input can achieve the maximum Degrees of Freedom (DoF) of the channel.

\item
In Section \ref{sec:achScheme} we evaluate the achievable rate region in \cite[Lemma 3]{oneObliviISIT2013} for the G-IC-OR by using a discrete input for the non-oblivious transmitter and a Gaussian input for the other transmitter. For simplicity we only consider the {\em symmetric} G-IC-OR, where the direct links have the same strength and the interfering links have the same strength, but our results can be readily extended to the general asymmetric case.

\item 
In past work on networks with oblivious nodes 
no performance guarantees were provided for the Gaussian noise case.
In Section\ref{sec:gDoFach} we study the generalized degrees of freedom (gDoF)
achievable with the scheme introduced in Section \ref{sec:achScheme}.
We show that in strong and very strong interference the {proposed scheme can approach the gDoF of the classical G-IC to within any degree of accuracy}.
This is quite surprising considering that the oblivious receiver can not perform joint decoding of the two messages, which is optimal for the classical G-IC in these regimes.

\item 
In Section \ref{sec:finiteSNR} we
show that the sum-capacity of the G-IC-OR in strong and very strong interference is within $O \left ( \log \left (\log (\snr ) \right) \right)$  of the sum-capacity of the classical IC (which forms a natural outer bound to the oblivious channel, and where we are able to compute outer bounds). 
{This in turn refines the gDoF result of Section \ref{sec:achScheme} and shows that the scheme introduced in Section \ref{sec:achScheme} is indeed gDoF optimal.}

\end{enumerate}

We conclude the paper with some final remarks and future directions  in Section \ref{sec:Conclusion}.

\subsection{Notation} 
Lower case variables are instances of upper case random variables which take values in calligraphic alphabets.
We let $\delta(\cdot)$ denote the  Dirac delta function, and $|A|$ denote the cardinality of a set $A$.  
The probability density function of a real-valued Gaussian random variable (r.v.) $X$ with mean $\mu$ and variance $\sigma^2$ is denoted as
\[
X \sim \mathcal{N}(x;\mu,\sigma^2) := \frac{1}{\sqrt{2\pi\sigma^2}}\eu^{-\frac{(x-\mu)^2}{2\sigma^2}}.
\]
Throughout the paper $\log(\cdot)$ denotes logarithms in base 2 and $\ln(\cdot)$ in base $\eu$.
We let $[x]^{+} :=\max(x,0)$ and $\log^{+}(x) :=[\log(x)]^+$.
The functions $\mud(N,x)$ and $\mug(x)$, for $N\in\mathbb{N}$ and $x\in\mathbb{R}^+$, are defined as
\begin{align*}
\mud(N,x) &:=
\left[\log(N)-\frac{1}{2}\log\left(\frac{\eu}{2}\right)
-\log\left(1+(N-1)\eu^{-x} \right)\right]^+
\\
\mug(x)&:=\frac{1}{2}\log(1+x).
\end{align*}
In the following $\text{PAM}(N,d_{\rm min})$ denotes the uniform distribution over a zero-mean Pulse Amplitude Modulation (PAM) constellation with $N$ points and minimum distance $d_{\rm min}$  (and average energy $\mathcal{E} = d_{\rm min}^2\frac{N^2-1}{12}$). 

\section{Channel Model}
\label{sec:channelModel}

\begin{figure}
 \center
 \includegraphics[width=9cm]{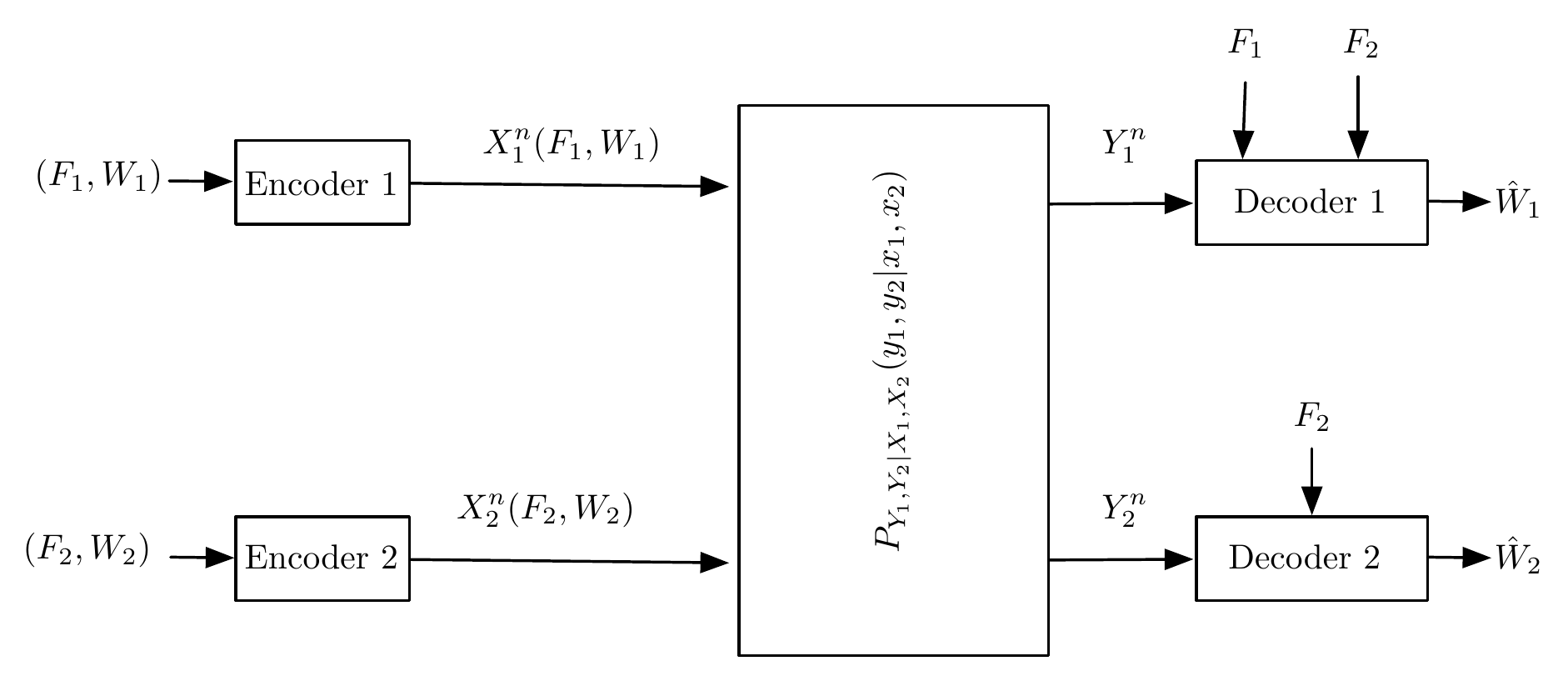}
 \caption{The IC-OR, where $F_1$ and $F_2$ represent codebook indices known to one or both receivers.}  \label{fig:channel:model}
\end{figure}

The IC-OR consists of a two-user memoryless IC $(\mathcal{X}_1,\mathcal{X}_2,P_{Y_1 Y_2|X_1 X_2},\mathcal{Y}_1,\mathcal{Y}_2)$ where receiver~2 is oblivious of  transmitter~1's codebook.  We model this lack of codebook knowledge as in \cite{sand_decentr_proces}, where transmitters use randomized encoding functions indexed by a message index and a {\em codebook index}. 
An oblivious receiver is unaware of the codebook index ($F_1$ is not given to decoder~2 in Fig. \ref{fig:channel:model}). The basic modeling assumption  is that without the knowledge of the codebook index a codeword looks unstructured. More formally, by extending \cite[Definition 2]{simion_codebook}, a ($2^{nR_1},2^{nR_2},n$) code for the IC-OR with time sharing is a six-tuple $(P_{F_1|Q^n},\sigma_1^n,\phi_1^n,P_{F_2|Q^n},\sigma_2^n,\phi_2^n)$, where the distribution $P_{F_i|Q^n}$, $i\in[1:2]$, is over a finite alphabet  $\mathcal{F}_i$ conditioned on the time-sharing sequences $q^n$ from some finite alphabet $\mathcal{Q}$, and where the encoders $\sigma_i^n$ and the decoders $\phi_i^n$, $i\in[1:2]$, are mappings 
\begin{align*}
&\sigma_1^n: [1:2^{nR_1}] \times [1:|\mathcal{F}_1|] \rightarrow \mathcal{X}_1^n,\\
&\sigma_2^n: [1:2^{nR_2}] \times [1:|\mathcal{F}_2|] \rightarrow \mathcal{X}_2^n,\\
&\phi_1^n: [1:|\mathcal{F}_1|]  \times [1:|\mathcal{F}_2|] \times \mathcal{Y}_1^n \rightarrow [1:2^{nR_1}],\\
&\phi_2^n: [1:|\mathcal{F}_2|] \times \mathcal{Y}_2^n \rightarrow [1:2^{nR_2}].
\end{align*}
Moreover, when transmitter 1's codebook index is unknown at decoder 2,  the encoder $\sigma_1^n$ and  distribution $P_{F_1|Q^n}$ satisfy 
\begin{align}
  &\sum_{w_1=1}^{2^{nR_1}} \sum_{f_1=1}^{|\mathcal{F}_1|}P_{F_1|Q^n}(f_1|q^n) \ 2^{-nR_1} \ \delta\big(x_1^n - \sigma_1^n(w_1,f_1)\big) \nonumber
\\&=: \mathbb{P}[X_1^n = x_1^n | Q^n = q^n]=\prod_{t\in[1:n]} P_{X_1|Q}(x_{it}|q_t)
\label{eq:osvaldo:eq},
\end{align}
according to some distribution $P_{X_1|Q}$. 
In other words, when averaged over the probability of selecting a given codebook and over a uniform distribution on the message set, the transmitted codeword conditioned on any time sharing sequence has a product distribution. 
Besides the restriction in~\eqref{eq:osvaldo:eq} on the allowed class of codes, the probability of error, achievable rates and capacity region are defined in the usual way~\cite{elgamalkimbook}. 

In this work we consider the practically relevant real-valued single-antenna symmetric Gaussian noise case.
The restriction to symmetric channel gains is just for ease of exposition; all the results in the following can be extended straightforwardly to the general asymmetric case.
For the symmetric G-IC-OR, the input-output relationship is
\begin{subequations}
\begin{align}
Y_1=\sqrt{\snr} \ X_1+\sqrt{\inr} \ X_2+Z_1\\
Y_2=\sqrt{\inr} \ X_1+\sqrt{\snr} \ X_2+Z_2 
\end{align}
\label{ch:gauss}
\end{subequations}
where the channel inputs  are subject to the average power constraint $\mathbb{E}[|X_{i}|^2] \le 1, i\in[1:2]$,
and the noise are i.i.d. $Z_i\sim \mathcal{N}(z; 0,1), i\in[1:2]$. 
The real-valued parameters $\snr$ and $\inr$ represent the received signal-to-noise ratio of the intended and interfering signal, respectively, at each receiver.

\section{Main Tool}
\label{sec:MainTool}
In this section we present a new lower bound on the mutual information achievable by a discrete input on a point-to-point Gaussian noise channel that will serve as the main tool in evaluating our inner bound for the G-IC-OR.
We are not the first to consider discrete inputs for Gaussian noise channels; however, to best of our knowledge, prior to this, no firm lower bounds existed.
\cite[Theorems 6 and 7]{WuVerduDOFAllerton10} asymptotically characterize the optimal input distribution over $N$ masses at high and low $\snr$, respectively, for a point-to-point power-constrained Gaussian noise channel; \cite[Theorem 8]{WuVerduDOFAllerton10} gives a mutual information lower bound that holds for the {\it Gauss quadrature} distribution for all $\snr$s;
\cite{Alvarado:highSNR} considers arbitrary input constellations with distribution {\it independent of $\snr$} and finds exact asymptotic expressions for the rate in the high-SNR limit. 
Here we can not use these results as we need {\it firm} lower bounds that hold for all distributions of $N$ distinct masses 
and {\it for all} $\snr$. Our bound is as follows.

\begin{thm} 
\label{thm:lowbound}
Let $X_D$ be a discrete random variable with support $\{s_i \in \mathbb{R}, i\in[1:N]\}$, minimum distance $d_{\rm min}$ and average energy $\mathcal{E}_D:= \sum_{i\in[1:N]} s_i^2 \mathbb{P}[X_D=s_i]$.  
Let $Z_G\sim \mathcal{N}(z; 0,1)$ and $\snr$ be a non-negative constant. 
Then
\begin{align}
 \mud\left(N, \snr \ \frac{d_{\rm min}^2}{4}\right)
 \label{eq:lower Xd}
 \leq I(X_D; \sqrt{\snr} \ X_D+Z_G) 
 \\ 
 \leq \min\Big(\log(N), \mug\left(\snr \ \mathcal{E}_D\right) \Big).
 \label{eq:upper Xd}
\end{align}
\end{thm}

\begin{IEEEproof}
Let $p_i:= \mathbb{P}[X_D=s_i], i\in[1:N]$.
The output $Y= \sqrt{\snr} \ X_D+Z_G$ has density
\begin{align}
Y \sim  P_{Y}(y) := \sum_{i\in[1:N]} p_i\mathcal{N}(y; \sqrt{\snr} \  s_i,1).
\label{eq:gauss mixture PY}
\end{align}
The upper bound in~\eqref{eq:upper Xd} follows from the well known facts that `Gaussian maximizes the differential entropy for a given second moment constraint' and that `a uniform input maximizes the entropy of a discrete random variable' \cite{elgamalkimbook}.
To prove the lower bound in~\eqref{eq:lower Xd} we first find a lower bound on the differential entropy
$
h(Y):=-\int P_{Y}(y) \log( P_{Y}(y) ) dy,
$ 
where the output density is the Gaussian mixture in~\eqref{eq:gauss mixture PY}.
We have
\begin{align*}
&-h(Y)=\int P_{Y}(y) \log( P_{Y}(y) ) dy\\
&\stackrel{(a)}{ \le}\log \int P_{Y}(y) P_{Y}(y) dy\\
&  = \log \int \left( \sum_{i\in[1:N]} p_i\mathcal{N}(y; \sqrt{\snr}   s_i,1) \right)^2 dy\\
&  = \log  \Big( \sum_{(i,j)\in[1:N]^2} p_i p_j \ \int \mathcal{N}(y; \sqrt{\snr}  s_i,1)  
\\& \quad \cdot \mathcal{N}(y; \sqrt{\snr}   s_j,1) dy \Big) \\
&  = \log  \Big( \sum_{(i,j)\in[1:N]^2} p_i p_j \frac{1}{\sqrt{4\pi}} \eu^{\frac{-\snr (s_i-s_j)^2}{4}}
\\& \quad \cdot \int \mathcal{N}(y;  \sqrt{\snr}  \frac{s_i+s_j}{2},\frac{1}{2})  dy \Big) \\
&  \stackrel{(b)}{=} \log  \left( \sum_{(i,j)\in[1:N]^2} p_i p_j \frac{1}{\sqrt{4\pi}} \eu^{-\frac{\snr(s_i-s_j)^2}{4}}\right)  \\
& \stackrel{(c)}{\le} \log  \Big( 
 \sum_{i\in[1:N]} p_i^2 \frac{1}{\sqrt{4\pi}}
+\sum_{i\in[1:N]} p_i(1-p_i) \frac{1}{\sqrt{4\pi}} \eu^{-\frac{ \snr d_{\rm min}^2}{4}}\Big) \\
& \stackrel{(d)}{\le} -\log(N\sqrt{4\pi}) + \log  \left( 
 1 + (N-1) \eu^{-\frac{ \snr d_{\rm min}^2}{4}}\right) ,
\\&\Longleftrightarrow I(X_D; \sqrt{\snr} \ X_D+Z_G)=h(Y)- h(Z_G)\geq 
\\&\log(N)-\frac{1}{2}\log\left(\frac{\eu}{2}\right) - \log  \left( 
 1 + (N-1) \eu^{-\frac{ \snr d_{\rm min}^2}{4}}\right),
\end{align*}
where the (in)equalities follow from:
(a) Jensen's inequality, 
(b) $\int \mathcal{N}(y;  \mu,\sigma^2)  dy=1$, 
(c) upper bounding by maximizing the exponential with  $d_{\rm min}:= \min_{i\not=j}|s_i-s_j|$, 
(d) by maximizing over the $\{p_i, i\in[1:N]\}$. 
Combining this bound with the fact that mutual information is non-negative proves the lower bound in~\eqref{eq:lower Xd}.
\end{IEEEproof}

\section{Discrete inputs for the power-constrained point-to-point Gaussian noise channel}
\label{sec:ptpExample}

In this section we 
give a flavor of how we intend to use discrete inputs on the G-IC-OR by considering the familiar point-to-point 
Gaussian noise channel.
Specifically, we will show that, for a unit-variance additive white Gaussian noise channel, 
the unit-energy discrete input $X_D \sim \text{PAM}\left(N,\sqrt{\frac{12}{N^2-1}}\right)$ with a properly chosen number of points $N$ as a function of $\snr:=|h|^2$ achieves
\begin{align}
I(X_D;hX_D+Z_G)     & \approx \log(N), \label{eq:discrete:input}
\\
I(X_G;hX_G+X_D+Z_G) & \approx I(X_G;hX_G+Z_G), \label{eq:discrete:interf}
\end{align}
What this implies is that the discrete input $X_D$ is a ``good'' input and a ``good'' interference. To put it more clearly, when we use a discrete constellation with uniform distribution as input, as in \eqref{eq:discrete:input}, the mutual information is roughly equal to the entropy of the constellation, which is highly desirable. On the other hand, when the same constellation is used as interference/noise, as in \eqref{eq:discrete:interf}, the mutual information is roughly as if there was no interference, which is again  highly desirable.  In contrast, a Gaussian r.v. is considered to be the ``best'' input but the ``worst'' interference/noise when subject to a second moment constraint \cite{WANlemma}. 


Consider the point-to-point Gaussian channel
\begin{subequations}
\begin{align}
 Y=\sqrt{\snr}\ X+Z,
 \\ \mathbb{E}[X^2]\leq 1, \ Z\sim \mathcal{N}(z; 0,1),
\end{align}
\label{ptp:InputOutput}
\end{subequations}
whose capacity $C=\mug\left(\snr\right)$ is
achieved by $X  \sim \mathcal{N}(x; 0,1)$ at all $\snr$s.
For this channel the gDoF is
\begin{align}
d:=\lim_{\snr \to \infty} \frac{C}{\frac{1}{2} \log(1+\snr)}=1. 
\label{eq:gDoF:def}
\end{align}

Consider now the performance of the input
\begin{align}
 X  \sim \text{PAM}\left(N,\sqrt{\frac{12}{N^2-1}}\right).
\label{eq:ptp input N}
\end{align}
It was shown in \cite[Th. 10]{MMSEdim} that 
for any fixed $N$ {\em independent} of $\snr$ the gDoF is zero.
Similar conclusions were found in \cite{WuVerduDOFAllerton10} by considering high-SNR approximations of the {\em finite constellation} capacity of the point-to-point Gaussian channel, defined as the maximum rate achieved by a discrete input constrained to have a finite support. A question left open in \cite{WuVerduDOFAllerton10} is what happens if $N$ is allowed to be a function of $\snr$. In the following we address this question.

\begin{figure}
\center
\includegraphics[width=8cm]{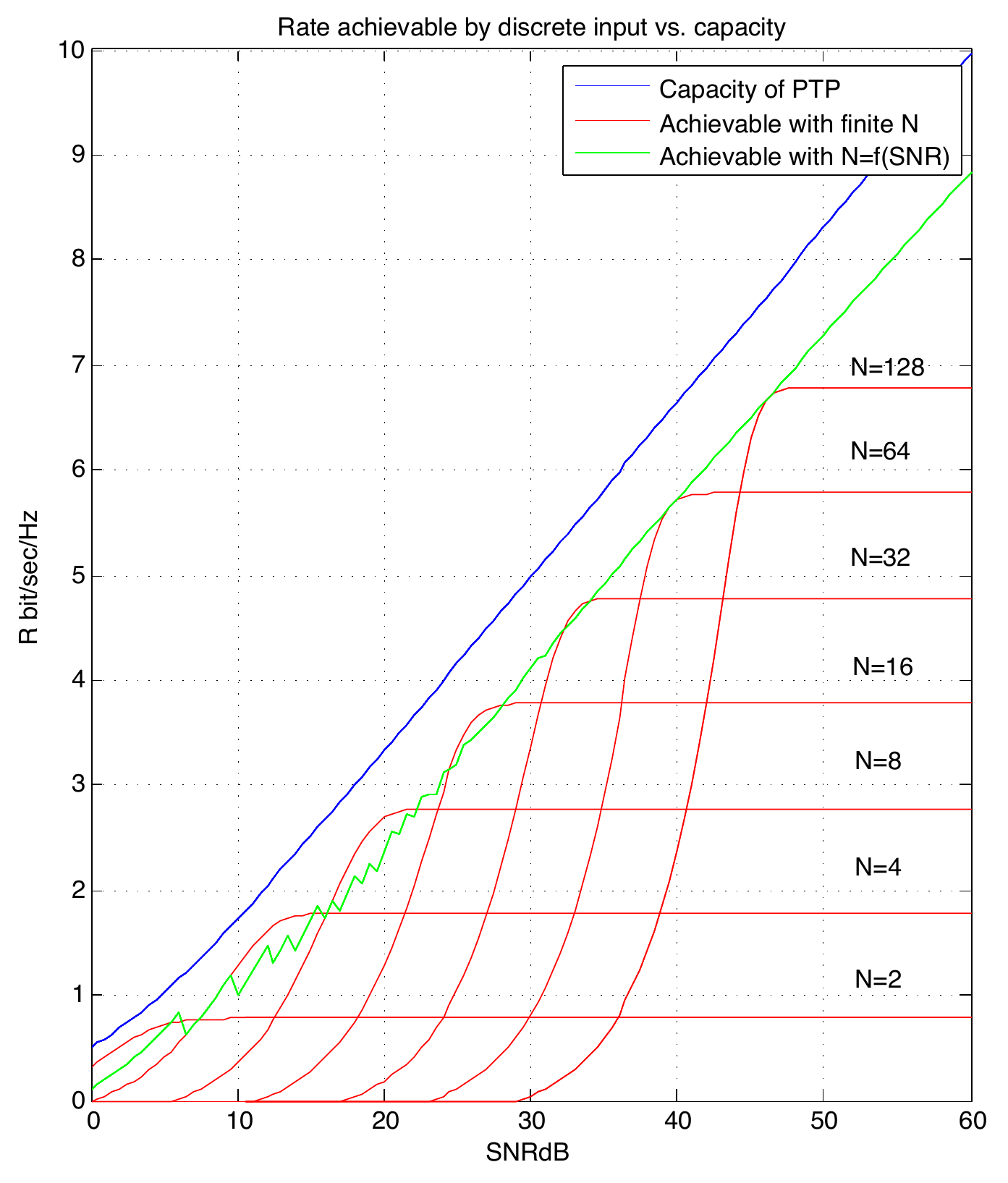}
\caption{Comparing achievable strategies: 1) Gaussian input (blue line), 2) Discrete input with fixed $N$ (red lines), 3) Discrete input with $N$ as in Theorem ~\ref{thm:ptp:finiteS} (green line).}
\label{fig:ptpAch}
\end{figure}

Fig.~\ref{fig:ptpAch} shows that, through clever picking of $N$ as a function of $\snr$, we seem to be able to follow to within an additive gap the capacity $C=\mug\left(\snr\right)$.
This is formally shown in the next theorem.
\begin{thm}
\label{thm:ptp:Th gdof}
For the channel in~\eqref{ptp:InputOutput}, the input in~\eqref{eq:ptp input N}  with
\begin{align}
 N=\lfloor \sqrt{1+\snr^{1-\epsilon}} \rfloor
\label{eq:ptp input choice of N}
\end{align}
achieves $d=1-\epsilon$, for any $\epsilon \, { \in (0,1)}$.
\end{thm}
\begin{IEEEproof}
By Theorem~\ref{thm:lowbound}, the proposed input achieves
\begin{align}
R \geq 
\mud\left(\lfloor \sqrt{1+\snr^{1-\epsilon}}\rfloor, \  3 \snr^{\epsilon}\right)
\label{eq:ptp rate with input choice of N}
\end{align}
Next, by using the definition of gDoF and the fact that
\begin{align*}
\frac{\log\left(1+(\lfloor \sqrt{1+\snr^{1-\epsilon}} \rfloor-1)\eu^{-\frac{3 \snr }{\lfloor \sqrt{1+\snr^{1-\epsilon}} \rfloor^2-1}} \right)}{\frac{1}{2}\log(1+\snr)} \to 0,
\end{align*}
for any $\epsilon>0$, we see that
\begin{align*}
d=\lim_{\snr \to \infty} \frac{\log(N)}{\frac{1}{2}\log(1+\snr)} = 1-\epsilon
\end{align*}
as claimed. This concludes the proof.
\end{IEEEproof}

Theorem~\ref{thm:ptp:Th gdof} shows that the input in~\eqref{eq:ptp input N} with the number of points chosen as in~\eqref{eq:ptp input choice of N} approaches 1~gDoF to within any degree of accuracy. We next show that, with a clever choice of $\epsilon$ in~\eqref{eq:ptp input choice of N} as a function of $\snr$, the rate in~\eqref{eq:ptp rate with input choice of N} is to within an additive gap of $O(\log(\log(\snr)))$ of the capacity $C=\mug\left(\snr\right)$, thus showing that indeed a discrete input can {\em exactly achieve} 1~gDoF. 
As a matter of fact this gap (given precisely in~\eqref{eq:ptp final upper on gap}) grows very slowly with $\snr$. For example, the gap reaches value of 3 bits at $\snr \approx  50  \text{dB}$. Hence, for all practical purposes, this gap can be considered a small constant. 
We have
\begin{thm}
\label{thm:ptp:finiteS}
For the channel in~\eqref{ptp:InputOutput}, the input in~\eqref{eq:ptp input N} with the number of points chosen as in~\eqref{eq:ptp input choice of N} and with
\begin{align}
\epsilon \, { =} \left[\frac{\log(\frac{1}{6} \ln(\snr))}{\log(\snr)} \right]^{+}
\label{eq:ptp input choice of epsilon}
\end{align}
the achievable rate in~\eqref{eq:ptp rate with input choice of N} achieves the capacity $C=\mug\left(\snr\right)$ to within an additive gap of $O(\log(\log(\snr)))$.
\end{thm}
\begin{IEEEproof}
Picking $\epsilon$  as in~\eqref{eq:ptp input choice of epsilon}
ensures that 
\begin{align*}
 \log\left(1+(\lfloor \sqrt{1+\snr^{1-\epsilon}} \rfloor-1)\eu^{-\frac{3 \snr }{\lfloor \sqrt{1+\snr^{1-\epsilon}} \rfloor^2-1}} \right) \le 1,
\end{align*}
and hence the achievable rate satisfies
\begin{align}
R \geq \left[\log(N)-\frac{1}{2}\log\left(\frac{\eu}{2}\right) -1  \right]^+.
\label{eq:simp:rate}
\end{align}
Next, the difference between the capacity and the achievable rate in \eqref{eq:simp:rate} for $\snr \ge 1$ (if $\snr<1$ a trivial gap of 1 bit/sec/Hz can be shown) can be upper bounded as
\begin{align}
&\frac{1}{2}\log(1+\snr) - \left[\log(\lfloor \sqrt{1+\snr^{1-\epsilon}} \rfloor)-\frac{1}{2}\log\left(\frac{\eu}{2}\right) -1  \right]^+ 
\notag\\
&\le \frac{1}{2}\log(1+\snr) - \log(\lfloor \sqrt{1+\snr^{1-\epsilon}} \rfloor)+\frac{1}{2}\log\left(\frac{\eu}{2}\right) +1 
\notag \\
&\le \frac{1}{2}\log(1+\snr) - \frac{1}{2} \log({1+\snr^{1-\epsilon}} )+\frac{1}{2}\log\left(\frac{\eu}{2}\right) +2
\notag
\end{align}
where used $\lfloor x \rfloor \ge \frac{1}{2}x$ for $x\geq 1$; next, since 
$\frac{1+x}{1+x^{1-\epsilon}} \le x^\epsilon$ for $x \geq 1$ (for $\snr \leq 1$ capacity can be trivially achievable to within 1~bit)
we have 
\begin{align}
\text{gap}(\snr)&\le  \frac{\epsilon}{2}\log(\snr) + { \frac{1}{2}\log\left(8\eu\right)} \notag\\
&=\left[{\frac{1}{2}\log(\frac{1}{6} \ln(\snr))} \right]^{+}+{ \frac{1}{2}\log\left(8\eu\right)},\label{eq:ptp final upper on gap}
\end{align}
as claimed. This concludes the proof. 
\end{IEEEproof}

\medskip
Theorem~\ref{thm:ptp:finiteS} showed that a discrete input is a ``good'' input in the sense alluded to by~\eqref{eq:discrete:input}. We now show that that a discrete interference is a ``good'' interference in the sense alluded to by~\eqref{eq:discrete:interf}. We study an extension of the channel in~\eqref{ptp:InputOutput} by considering a {\em state $T$ available neither at the encoder nor at the decoder}. The input-output relationship is
\begin{subequations}
\begin{align}
 Y=\sqrt{\snr}\ X+hT+Z :
\\ \mathbb{E}[X^2]\leq 1, \ Z\sim \mathcal{N}(z; 0,1),
\\ T\sim \text{PAM}\left(N,\sqrt{\frac{12}{N^2-1}}\right).
\end{align}
\label{ptp:InputOutput state}
\end{subequations}
It is well known \cite{elgamalkimbook} that capacity of channel with random state is 
$C=\max_{P_X} I(X;Y) \leq \max_{P_X} I(X;Y|T)= \mug(\snr)$.
We can show 
\begin{thm}
\label{thm:ach:ptp:state}
For the channel with unknown states in~\eqref{ptp:InputOutput state} the input $X \sim  \mathcal{N}(x;0,1)$ achieves
\begin{align}
R &{ \geq} \mug(\snr)+\mud\left(N, \frac{3|h|^2}{(1+\snr) (N^2-1)}\right)
\notag\\&-\min \left(\log(N),\mug(|h|^2)\right). 
\label{eq:thm:ach:ptp:state}
\end{align}
\end{thm}
\begin{IEEEproof}
By using Theorem~\ref{thm:lowbound} we have
\begin{align*}
  &I(X;Y)=
h(\sqrt{\snr}X+hT+Z) -h(hT+Z)
\\&=h\left(\frac{h}{\sqrt{1+\snr}}T+Z\right)-h(Z)+\frac{1}{2}\log(1+\snr)
\\&-\Big(h\left(h T+Z\right)-h(Z)\Big)
\\&\geq \mud\left(N, \frac{3|h|^2}{(1+\snr) (N^2-1)}\right)+\mug(\snr)
\\&-\min\Big(\log(N),\mug(|h|^2)\Big),
\end{align*}
as claimed.
\end{IEEEproof}
Note that the result of Theorem~\ref{thm:ach:ptp:state} can be readily used to lower bound the achievable rate in a G-IC where one user has a Gaussian input and the other a discrete input and where the discrete input is `treated as noise', as we shall do in the next Section for the G-IC-OR.
%
Before concluding we show that
\begin{cor}
\label{thm:ach:ptp:state gdof}
For the channel with unknown states in~\eqref{ptp:InputOutput state}, the achievable rate from Theorem~\ref{thm:ach:ptp:state} attains 1~gDoF.
\end{cor}
\begin{IEEEproof}
By taking achievable rate in Theorem~\ref{thm:ach:ptp:state} we have $d=1$
since both $\mud\left(N, \frac{3|h|^2}{(1+\snr) (N^2-1)}\right)$ and $\min\Big(\log(N),\mug(|h|^2)\Big)$ tend to zero as $\snr \to \infty$ (here $N$ and $|h|$ do not depend on $\snr$).
\end{IEEEproof}
Theorem~\ref{thm:ach:ptp:state gdof} shows that even with lack of state knowledge at both the receiver and transmitter, if the state is discrete and {\em its support is not a function of $\snr$,} then its effect can be `removed' at high-SNR. {From the proof of Theorem~\ref{thm:ach:ptp:state gdof} it is immediate that  the channel with unknown states in~\eqref{ptp:InputOutput state} has 1~DoF also when $N$ and $|h|$ vary with $\snr$ as long as $\mud\left(N, \frac{3|h|^2}{(1+\snr) (N^2-1)}\right) -\min\Big(\log(N),\mug(|h|^2)\Big)$ tends to zero as $\snr \to \infty$ in the rate expression in~\eqref{eq:thm:ach:ptp:state}.}


\medskip
In the next Section we shall use a discrete input to characterize the sum-capacity of the G-IC-OR.

\section{An achievable region for the G-IC-OR}
\label{sec:achScheme}
With the tools and insights developed form the previous Sections, we are ready to analyze the G-IC-OR. 
\begin{thm}
\label{thm:sec:achScheme G-IC-OR}
For the G-IC-OR the following rate region is achievable
\begin{subequations}
\begin{align}
R_1 &\leq \mud\left(N, \frac{3 \ \snr}{N^2-1}\right)
\label{eq:ach.reg. Gaussian ICOR U2=X2 R1}
\\
R_2 & \leq \mud\left(N, \frac{3 \ \inr}{(1+\snr) (N^2-1)}\right)+\mug(\snr)
\notag\\&\qquad-\min \left(\log(N),\mug(\inr)\right)
\label{eq:ach.reg. Gaussian ICOR U2=X2 R2}
\\
R_1+R_2 &\leq\mud\left(N, \frac{3 \ \snr}{(1+\inr) (N^2-1)}\right)
+\mug(\inr)
\label{eq:ach.reg. Gaussian ICOR U2=X2 R1+R2}
\end{align}
\label{eq:ach.reg. Gaussian ICOR U2=X2}
\end{subequations}
\end{thm}
\begin{IEEEproof}
From \cite[Lemma 3 with $U_2=X_2$]{oneObliviISIT2013} the following 
region is achievable
%
\begin{subequations}
\begin{align}
R_1     & \leq I(X_1;Y_1|X_2,Q)\\
R_2     & \leq I(X_2;Y_2|Q)\\
R_1+R_2 & \leq I(X_1,X_2;Y_1|Q),
\end{align}
for all $P_QP_{X_1|Q}P_{X_2|Q}$.
\label{eq:ach.reg. ICOR U2=X2}
\end{subequations}
We now evaluate the region in~\eqref{eq:ach.reg. ICOR U2=X2} without time sharing, i.e., $Q=\emptyset$, and with inputs 
\begin{subequations}
\begin{align}
X_1& \sim \text{PAM}\left(N,\sqrt{\frac{12}{N^2-1}}\right),
\\
X_2& \sim {\mathcal{N}(x;0,1)}.
\end{align}
\label{eq:inputs}
\end{subequations}
The bound in~\eqref{eq:ach.reg. Gaussian ICOR U2=X2 R1} is a direct application of Theorem~\ref{thm:lowbound}
with $d_{\rm min}^2=\frac{12}{N^2-1}$. The bound in~\eqref{eq:ach.reg. Gaussian ICOR U2=X2 R2} follows from  Theorem~\ref{thm:ach:ptp:state} with $|h|^2=\inr$. The bound in~\eqref{eq:ach.reg. Gaussian ICOR U2=X2 R1+R2} follows since $I(X_1,X_2;Y_1) = I(X_1;Y_1)+I(X_2;Y_1|X_1)$, where $I(X_1;Y_1)$ is evaluated with Theorem~\ref{thm:lowbound} (here the Gaussian input $X_2$ is treated as noise hence the SNR is $\frac{\snr}{1+\inr}$; the minimum distance is $d_{\rm min}^2=\frac{12}{N^2-1}$) and $I(X_2;Y_1|X_1)=\mug(\inr)$.
\end{IEEEproof}

\section{High SNR performance}
\label{sec:gDoFach}
In this Section we analyze the performance of the scheme in Theorem~\ref{thm:sec:achScheme G-IC-OR} at high-SNR by using the gDoF region as metric. For each rate $R_i$ we define a gDoF $d_i$ as in \eqref{eq:gDoF:def}, for $i\in[1:2]$, where we parameterize $\inr=\snr^\alpha$ for some $\alpha \ge 0$ \cite{etkin_tse_wang}.
In a spirit of Theorem~\ref{thm:ptp:Th gdof}, we take 
$N= \lfloor \sqrt{1+ \snr^{\beta}}\rfloor$.
With this, we have that the following  achievable gDoF region
\begin{thm}
\label{thm:ach:gDoF:Region}
From Theorem~\ref{thm:sec:achScheme G-IC-OR}, the following $(d_1,d_2)$ pairs are achievable 
\begin{subequations}
\begin{align}
&d_1  \le \left\{\begin{array}{ll}
\beta & \text{if }1-\beta > 0    \\
0     & \text{if } 1-\beta \leq 0 \\
\end{array}\right\},
\label{eq:thm:ach:gDoF:Region d1}
\\
&d_2 \le \left\{\begin{array}{ll}
\beta & \text{if } [\alpha-1]^{+}-\beta > 0    \\
0     &[ \text{if } \alpha-1]^{+}-\beta \leq 0 \\
\end{array}\right\} +1 -\min(\beta,\alpha),
\label{eq:thm:ach:gDoF:Region d2}
\\
&d_1+d_2   \le \left\{\begin{array}{ll}
\beta & \text{if } [1-\alpha]^{+}-\beta > 0    \\
0        & \text{if } [1-\alpha]^{+}-\beta \leq 0 \\
\end{array}\right\}+\alpha.
\label{eq:thm:ach:gDoF:Region d1+d2}
\end{align}
union over all $\beta\geq 0$.
\label{eq:thm:ach:gDoF:Region}
\end{subequations}
\end{thm}
\begin{IEEEproof}
Due to the space limitations we show the proof for $d_2$ in~\eqref{eq:thm:ach:gDoF:Region d2} only;
proofs for the other constraints follow similarly. 
By using $\inr=\snr^\alpha$ and $N= \lfloor \sqrt{1+ \snr^{\beta}}\rfloor$ and by noting that
\[
\lim_{\snr \to \infty}\frac{\log(N^2)}{\log(1+\snr)}    = \beta, \ 
\lim_{\snr \to \infty}\frac{\log(1+\inr)}{\log(1+\snr)} = \alpha,
\]
we compute $d_2$ in the following way
 \begin{align*}
 d_2&=\lim_{\snr \to \infty} \frac{\text{left hand side of eq.\eqref{eq:ach.reg. Gaussian ICOR U2=X2 R2}}}{\frac{1}{2}\log(1+\snr)}\\
 &=\beta -\lim_{\snr \to \infty}\frac{ \log\left( 1 + \left(N-1\right) \exp \left({ \frac{- 3\inr}{{(1+\snr)}(N^2-1)} } \right) \right)}{\frac{1}{2}\log(\snr)}\\
 &+1-\min(\beta,\alpha)\\
 &=\beta -\left\{\begin{array}{ll}
0&\alpha-1-\beta > 0    \\
 \beta       &\alpha- 1-\beta \leq 0 \\
\end{array}\right\}-\min(\beta,\alpha)+1.
\end{align*}
This concludes the proof.
\end{IEEEproof}

\begin{figure}
\begin{center}
\includegraphics[width=8cm]{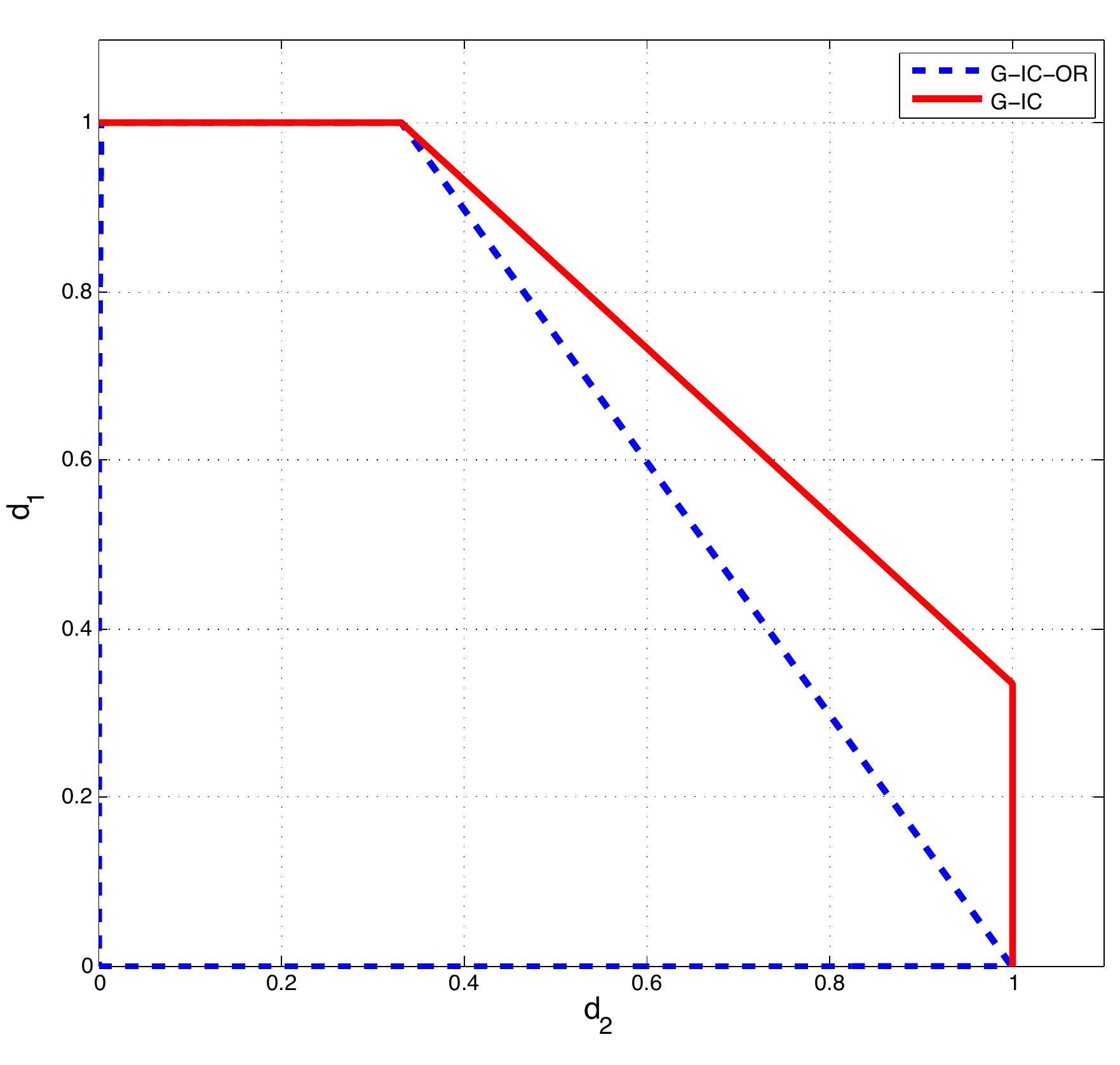}
\end{center}
\caption{gDof region achievable by Theorem~\ref{thm:ach:gDoF:Region} at $\alpha=\frac{4}{3}$.  { Note that one of the corner points is not achieved, as might be expected  since $d_2$ corresponds to the oblivious user.}  }
\label{fig:gdofexample}
\end{figure}

Determining analytically which $\beta$'s attain the closure of the gDoF region in~\eqref{eq:thm:ach:gDoF:Region} is a bit involved, but it can be done very easily numerally. Fig~\ref{fig:gdofexample} shows (for example) the gDoF region of Theorem~\ref{thm:ach:gDoF:Region} for $\alpha = \frac{4}{3}$; {we see that while the sum-gDoF is the same as that of the classical G-IC, that the region is not, which makes intuitive sense as $d_2$ corresponds to the achieved gDoF of the oblivious receiver, which is much more constrained in our model and our achievability scheme.} 

It is interesting to compare the achievable sum-gDoF for the G-IC-OR based on Thereom~\ref{thm:ach:gDoF:Region} with the outer bound given by sum-gDoF of the classical G-IC \cite{etkin_tse_wang}
\begin{align*}
\frac{\max(d_1+d_2)}{2} \leq \min \left (1,\max\left(\frac{\alpha}{2} , 1-\frac{\alpha}{2}\right),\max(\alpha,1-\alpha) \right).
\end{align*}
We next demonstrates the sum-gDoF of the G-IC-OR.
\begin{lem}
\label{lem:ach:gDoF}
The following gDoF is achievable by G-IC-OR 
\begin{align}
\max_{\beta}(d_1+d_2) = \left\{\begin{array}{ll}
{ 1-\epsilon}  & 0\leq \alpha <1, \ { 0< \epsilon \leq 1-\alpha} \\
{\alpha-\epsilon}  & 1\leq \alpha <2, \ {0< \epsilon \leq \alpha-1} \\
{2-\epsilon}  & \alpha \ge 2,    \ { 0< \epsilon \leq 1} \\
\end{array}\right..
\end{align}
\end{lem}
\begin{IEEEproof}
By setting
$\beta = \min(1,|\alpha-1|) - \epsilon \geq 0$ 
in Theorem~\ref{thm:ach:gDoF:Region} one can verify that 
\begin{align*}
   d_1     &\leq \beta,
\\ d_2     &\leq \min(1,\max(\alpha,1-\alpha)),
\\ d_1+d_2 &\leq \max(1,\alpha),
\end{align*}
from which the claim follows. 
\end{IEEEproof}

\begin{figure}
\center
\includegraphics[width=9cm]{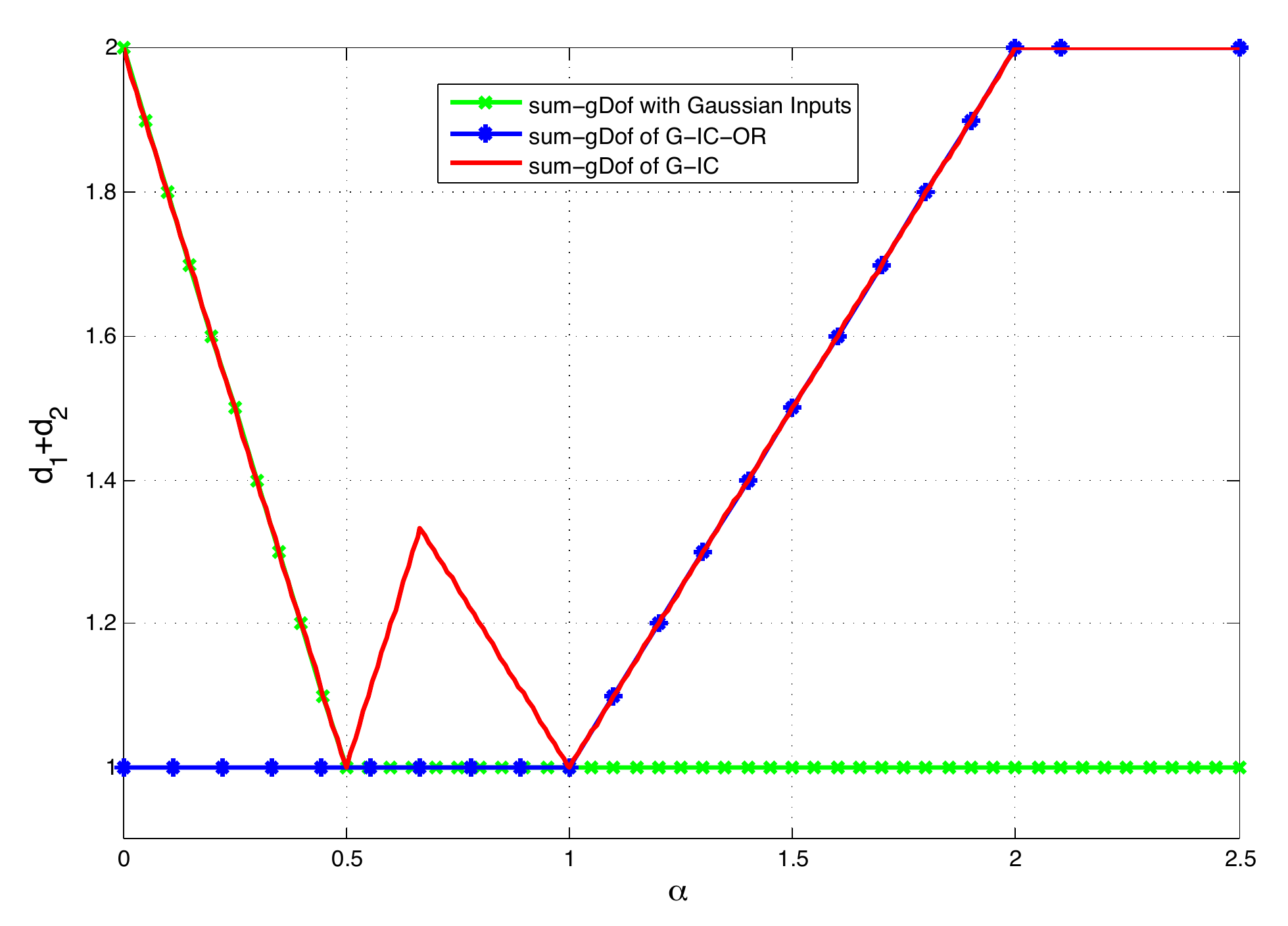}
\caption{The sum-gDoF of the G-IC-OR compared to the sum-gDoF of the G-IC. }
\label{fig:Wcurve}
\end{figure}

To compare performance of the classical G-IC and of the G-IC-OR we plot the corresponding gDoFs in Fig.~\ref{fig:Wcurve}. We observe that in strong and very strong interference ($\alpha\geq 1$) the gDoFs are the same (up to an arbitrary small $\epsilon$); hence, in this regime lack of codebook knowledge does not impact performance in the gDoF sense. {We also note that in weak interference ($\alpha< 1$) our proposed scheme only has 1~gDoF, which means that the same performance can be achieved by silencing one of the users. For reference we also plot the achievable sum-gDoF when both users use a Gaussian input and treat interference as noise, i.e., $d_1=d_2= [1-\alpha]^+$, which are known to be optimal for the classical G-IC in very weak interference ($\alpha\leq 1/2$) \cite{etkin_tse_wang}; we see that the proposed scheme does not achieve this sum-gDoF in this regime; indeed, in weak interference it is not optimal to set $U_2=X_2$ in \cite[Lemma 3]{oneObliviISIT2013} as we did in Theorem~\ref{thm:sec:achScheme G-IC-OR}; different choices of discrete inputs for the general region in \cite[Lemma 3]{oneObliviISIT2013} are reported in \cite{Dytso:ISIT2014}.}

\section{Finite SNR performance}

\label{sec:finiteSNR}
In the previous section we showed that in strong interference ($\alpha \ge 1$) the sum-gDoF of the classical G-IC can be approached with any precision even when one receiver lacks knowledge of the interfering codebook. Thus, it is interesting to ask whether one can exactly achieve  the sum-gDoF of the classical G-IC in strong interference by showing an additive gap to the  capacity of the  classical G-IC which is  $o(\log(\snr))$. In light of the results for the point-to-point channel, we ask whether we the sum-capacity of the G-IC-OR is to within  $O(\log(\log(\snr)))$  of that of the classical G-IC. We next answer this in the positive, thus showing that in strong interference there is no  penalty in term of sum-gDoF when one receiver is oblivious.

\begin{thm}
\label{gap:strong}
The sum-capacity of the G-IC-OR in the strong and very strong interference regimes
is to within $O(\log(\log(\snr)))$ of that of the classical G-IC.
\end{thm}
\begin{IEEEproof}
In very strong interference, $\snr(1+{\snr})\leq\inr$, we choose
\begin{align*}
N =\lfloor \sqrt{1+\snr^{1-\epsilon}} \rfloor, 
\
\epsilon =\left[ \frac{\log \left(\frac{1}{6}\ln(\snr)\right)}{\log(\snr)} \right ]^{+},
\end{align*}
for $N \ge 3$ in Theorem~\ref{thm:ach:gDoF:Region} so that the following rates are achievable for the G-IC-OR
\begin{align*}
R_1 &\le \log(\lfloor \sqrt{1+\snr^{1-\epsilon}} \rfloor)-\frac{1}{2}\log\left(\frac{\eu}{2}\right) -1,\\
R_2 &\le \frac{1}{2} \log(1+\snr)-\frac{1}{2}\log\left(\frac{\eu}{2}\right) -1.
\end{align*}
In this regime, the classical G-IC has capacity \cite{CarliealVeryStrongIntCap}
\begin{align*}
R_1 &\le \frac{1}{2} \log(1+\snr),\\
R_2 &\le \frac{1}{2} \log(1+\snr).
\end{align*}
Clearly, the gap for $R_2$ is a constant (with respect to $(\snr,\inr)$) given by $\frac{1}{2}\log\left(\frac{\eu}{2}\right)+1=1.2213$; the gap for $R_1$ is as in~\eqref{eq:ptp final upper on gap}. Although the theorem statement is for the sum-capacity, the proof holds for the whole capacity region. 

In strong interference, $\snr\leq\inr < \snr(1+{\snr})$, we choose 
\begin{align*}
N  = \left\lfloor \sqrt{1+\left(\frac{\inr}{1+\snr}\right)^{1-\epsilon} } \right\rfloor, 
\
\epsilon = \left[\frac{\log(\frac{1}{ 6} \ln(\frac{\inr}{1+\snr}))}{\log(\frac{\inr}{1+\snr})} \right]^{+},
\end{align*}
for $N \ge 3$ in Theorem~\ref{thm:ach:gDoF:Region} so that the following sum-rate is achievable for the G-IC-OR
\begin{align*}
R_1+R_2 &\le \log \left( \left\lfloor \sqrt{1+\left(\frac{\inr}{1+\snr}\right)^{1-\epsilon} } \right\rfloor \right)\\& + \frac{1}{2} \log(1+\snr)-\log \left(\frac{\eu}{2}\right) -2.
\end{align*}
In this regime, the classical G-IC has sum-capacity \cite{sato_strong} 
\begin{align*}
R_1+R_2 \le \frac{1}{2}\log(1+\snr+\inr).
\end{align*}
The gap is hence
\begin{align*}
&\frac{1}{2} \log(1+\snr+\inr)-  \log\left( \left \lfloor\sqrt{1+\left(\frac{\inr}{1+\snr}\right)^{1-\epsilon}} \right\rfloor\right)\\ &-\frac{1}{2} \log(1+{\snr})
+2+\log\left(\frac{\eu}{2}\right)
\\&\leq \left[{\frac{1}{2}\log\left(\frac{1}{6} \ln\left(\frac{\inr}{1+\snr}\right)\right)} \right]^{+}+{ 3+\log\left(\frac{\eu}{2}\right)}\\
&\leq \left[{\frac{1}{2}\log\left(\frac{1}{6} \ln\left(\snr\right)\right)} \right]^{+}+{ \log\left(4\eu\right)},
\end{align*}
since the steps are the same as those leading to~\eqref{eq:ptp final upper on gap} if one substitutes $\snr$ in~\eqref{eq:ptp final upper on gap} with $\frac{\inr}{1+\snr}$. Since $\frac{\inr}{1+\snr}\leq \snr$ in strong interference, we obtain an $O(\log(\log(\snr))$ gap in this regimeas well. 
This concludes the proof. 
\end{IEEEproof}
 
\section{Conclusion} 
\label{sec:Conclusion}
In the paper we focused on deriving capacity results for the Gaussian interference channel where one of the receivers is lacking knowledge of the interfering codebook, in contrast to a classical model where both receivers possess full codebook knowledge.
To that end we derived a novel inequality on the achievable rate in a point-to-point Gaussian noise channel with discrete inputs,
that we believe might be of an interest on its own. We surprisingly demonstrated that lack of codebook knowledge is not as detrimental as one might believe in strong and very strong interference. 

\section*{Acknowledgment}
The work of the authors was partially funded by NSF under award 1017436.
The contents of this article are solely the responsibility of the authors and do not necessarily
represent the official views of the NSF.

\bibliography{refs}
\bibliographystyle{IEEEtran}
\end{document}